\documentclass[twocolumn,tighten]{aastex63}

\usepackage{times,natbib,graphicx,amsmath,multirow,xspace}
\usepackage{xcolor}
\usepackage{lineno}

\newcommand{\nustar}{NuSTAR\xspace}

\newcommand{\nicer}{NICER\xspace}

\newcommand{\lc}{light curve}

\newcommand{\lumcgs}{ergs~s$^{-1}$\xspace}
\newcommand{\rin}{$R_{\rm in}$\xspace}
\newcommand{\rg}{$R_{g}$\xspace}
\newcommand{\risco}{$R_{\mathrm{ISCO}}$\xspace}

\newcommand{\xillver}{{\sc xillver}\xspace}
\newcommand{\xillverco}{{\sc xillverCO}\xspace}
\newcommand{\source}{\mbox{4U~0614$+$091}\xspace}

\definecolor{pink}{RGB}{212,76,133}
\definecolor{sky}{RGB}{51,160,232}
\definecolor{lavender}{RGB}{170,120,240}

\shorttitle{NICER-NuSTAR View of 4U~0614+091}
\shortauthors{Moutard et al.}

\begin{document}

\title{Simultaneous NICER and NuSTAR Observations of the Ultra-compact X-ray Binary 4U 0614+091}

\correspondingauthor{D. L. Moutard}
\email{david.moutard@wayne.edu}

\author{D.~L.~Moutard}
\author[0000-0002-8961-939X]{R.~M.~Ludlam}
\affiliation{Department of Physics \& Astronomy, Wayne State University, 666 West Hancock Street, Detroit, MI 48201, USA}
\author[0000-0003-3828-2448]{J.~A.~Garc\'{i}a}
\affiliation{Cahill Center for Astronomy and Astrophysics, California Institute of Technology, 1200 E. California Blvd, MC 290-17, Pasadena, CA, 91125, USA}
\author[0000-0002-3422-0074]{D. ~Altamirano}
\affiliation{School of Physics and Astronomy, University of Southampton, University Rd, Southampton SO17 1BJ, UK.}
\author{D.~J.~K.~Buisson}
\affiliation{Independent Researcher}
\author[0000-0002-8294-9281]{E.~M.~Cackett}
\affiliation{Department of Physics \& Astronomy, Wayne State University, 666 West Hancock Street, Detroit, MI 48201, USA}
\author[0000-0002-4397-8370]{J.~Chenevez}
\affiliation{Department of Space Research \& Technology, Technical University of Denmark, Elektrovej, 327, 214, 2800 Kgs. Lyngby, Denmark}
\author[0000-0002-0092-3548]{N.~Degenaar}
\affiliation{Anton Pannekoek Institute for Astronomy, University of Amsterdam, Science Park 904, 1098 XH Amsterdam, Netherlands}
\author[0000-0002-9378-4072]{A.~C.~Fabian}
\affiliation{Institute of Astronomy, University of Cambridge, Madingley Rd, Cambridge CB3 0HA, UK}
\author[0000-0001-8371-2713]{J.~Homan}
\affiliation{Eureka Scientific, 2452 Delmer Street Suite 100, Oakland, CA 94602, USA}
\author[0000-0002-3850-6651]{A.~Jaodand}
\affiliation{Cahill Center for Astronomy and Astrophysics, California Institute of Technology, 1200 E. California Blv, MC 290-17, Pasadena, CA, 91125, USA}
\author[0000-0002-8403-0041]{S.~N.~Pike}
\affiliation{Department of Astronomy and Astrophysics, University of California, San Diego, 9500 Gilman Dr, La Jolla, CA, 92093, USA}
\author[0000-0001-6715-0423]{A.~W.~Shaw}
\affiliation{Department of Physics and Astronomy, Butler University, Indianapolis, IN, 46208, USA}
\author[0000-0001-7681-5845]{T.~E.~Strohmayer}
\affiliation{NASA Goddard Space Flight Center, 8800 Greenbelt Road, Greenbelt, Maryland 20771, USA}
\author[0000-0001-5506-9855]{J.~A.~Tomsick}
\affiliation{Space Sciences Lab, University of California, Berkeley, 7 Gauss Way, Berkeley, CA, 94720, USA}
\author[0000-0003-0870-6465]{B.~M.~Coughenour}
\affiliation{Space Sciences Lab, University of California, Berkeley, 7 Gauss Way, Berkeley, CA, 94720, USA}

\begin{abstract}
We present the first joint NuSTAR and NICER observations of the ultra-compact X-ray binary (UCXB) 4U 0614+091. This source shows quasi-periodic flux variations on the timescale of $\sim$days. We use reflection modeling techniques to study various components of the accretion system as the flux varies. We find that the flux of the reflected emission and the thermal components representing the disk and the compact object trend closely with the overall flux. However, the flux of the power-law component representing the illuminating X-ray corona scales in the opposite direction, increasing as the total flux decreases. During the lowest flux observation, we see evidence of accretion disk truncation from roughly 6 gravitational radii to 11.5 gravitational radii. This is potentially analogous to the truncation seen in black hole low-mass X-ray binaries, which tends to occur during the low/hard state at sufficiently low Eddington ratios. 

\end{abstract}


\section{Introduction} \label{sec:intro}
Ultra compact X-ray binaries (UCXBs) are a class of low-mass X-ray binary (LMXB) which are distinguished by a significantly shorter orbital period ($<$80 minutes)compared to ``typical" LMXBs, which tend to have periods between a few hours to days. To achieve such a short orbital period, the compact object is likely accreting material from a degenerate companion such as a white dwarf (WD) or helium star \citep{nelson86, savonije86}. These objects are persistent gravitational wave sources that could be of interest to future multi-messenger missions (\citealt{chen20}).They are also, more generally, probes of WD physics, accretion physics, and the physics of compact objects. 

In LMXB systems, X-rays are believed to originate from near the compact object, from the closest accretion inflow. The source of high-energy non-thermal photons is thought to be an X-ray corona which is located very near the compact object itself \citep{syunyaev91}. This corona of hot, highly accelerated electrons Compton upscatter seed photons from the disk or boundary layer, producing hard X-ray emission \citep{ibragimov05}. These X-rays can be viewed directly by observers, but they can also be reprocessed by regions in the disk. In many LMXBs these reprocessed features commonly appear as an Fe K$\alpha$ feature around 6.4~keV (\citealt{fabian89}). Because of the oxygen-rich WD companion in some UCXBs, the accretion disks are often devoid of hydrogen and helium \citep{nelemans04,nelemans06}, while the reflection spectrum displays an \ion{O}{8} Ly$\alpha$ line around 0.67 keV \citep{christian94, juett01}. 

X-ray reflection modeling is a technique used to model the photons which are reprocessed by the disk. By combining a continuum model with an X-ray reflection model, we can study the properties of the accretion disk and the X-ray source. The continuum portion of the model accounts for direct emission from the disk, corona, and the compact object itself. The reflected emission is broadened by Doppler, special and general relativistic effects. Hence, the degree of broadening in the reflection spectrum is correlated to the proximity to the compact object; yielding measurements of the inner disk radius. This technique has been used to constrain accretion disk parameters for many LMXBs and a handful of UCXBs \citep{miller07, cackett08, cackett09b, cackett10, madej14, ludlam21}. In this paper, we use simultaneous data from both \nicer \citep{gendreau12} and \nustar \citep{harrison13}, and we model the reflection spectra and probe the accretion disk of the UCXB \source.

\source was originally detected as an X-ray burster in 1975 \citep{swank78}. The presence of Type-I X-ray bursts have identified the compact source as a neutron star \citep{brandt92}, and the presence of oxygen and neon features indicates that the donor source is likely a WD \citep{juett01,nelemans04}. The orbital period of the system is approximately 50 minutes (\citealt{shahbaz08}). Previous spectral studies have shown evidence of both \ion{O}{8} and Fe K reflection features in the system (\citealt{madej10, madej14, ludlam19a}), and found that the Fe abundance is subsolar. \cite{migliari10} combined spectral data ranging from radio to X-ray, providing the first detection of the radio counterpart and characterizing the jet in the system. Using XMM-Newton and \xillverco (a reflection table with abundances matching those of a WD), \cite{madej14} studied \source and found that \xillverco improved reflection fit statistics by $\sim16\%$ over previous reflection models, and verified the existence of absorption edges around 0.88 keV.

The long term light curve for this object is variable and displays quasi-periodic behavior, peaking every few days. By studying the spectral changes over various phases of this light curve, we can study which components drive the changing flux, and whether the location of the inner accretion disk radius varies in response. This paper is divided as follows: in the next section we discuss the observations and data reduction. In Section \ref{sec:methods} we discuss the modeling methods used, and share our results. In Section \ref{sec:discussion} we discuss the implications of these results, then finally in Section \ref{sec:conclusions} we summarize these results and conclude.

\begin{table}[t!]
\caption{\source Observation Information}
\label{tab:obs}
\begin{center}
\begin{tabular}{llcccc}
\hline

Obs. & Mission & Sequence ID & Obs.\ Start (UTC) & Exp. (ks) \\
\hline
1 & NuSTAR & 30702009002 & 2021-10-06 05:36:09 & $  28.7$\\
& NICER & 4701010101 & 2021-10-06 06:05:29 & $  15.3$ \\

2 & NuSTAR & 30702009004 & 2021-10-09 09:16:09 & $  29.4$\\
& NICER & 4701010201 & 2021-10-09 08:06:17 & $  21.1$ \\
&  & 4701010202 & 2021-10-09 23:37:04 & $  2.1$ \\

3 & NuSTAR & 30702009006 & 2021-10-11 17:41:09 & $  29.0$\\
& NICER & 4701010301 & 2021-10-11 17:27:39 & $  7.6$ \\
&  & 4701010302 & 2021-10-11 23:35:00 & $  11.6$ \\

4 & NuSTAR & 30702009008 & 2021-10-13 17:56:09 & $  28.6$\\

5 & NuSTAR & 30702009010 & 2022-01-19 06:51:09 & $  28.2$\\
& NICER & 4701010401 & 2022-01-19 07:07:18 & $  10.4$ \\
\hline

\end{tabular}
\end{center}
\end{table}

\section{Observations and Data Reduction} \label{sec:data}
The source was observed on five occasions with \nustar, four of which were performed simultaneously with \nicer.
Table~\ref{tab:obs} provides the observing details for the contemporaneous \nicer and \nustar observations.
The \nustar data were reduced using the standard data reduction process with {\sc nustardas} v2.1.2 and {\sc caldb} 20221115.
Spectra and \lc s were extracted using a circular region with a 160$''$ diameter centered on the source. Backgrounds were generated from a 160$''$ diameter region on the same detector but away from the source.
The \nicer observations were reduced using {\sc nicerdas} 2022-10-20\_V010. Data were re-calibrated with the latest calibration files available in CALDB release 20221001 through implementation of the {\sc nicerl2} command. 
Good time intervals (GTIs) were generated using {\sc nimaketime} to select events that occurred when the particle background was low (KP~$<$~5 and COR\_SAX~$>$~4) and avoiding times of extreme optical light loading. The GTIs were applied to the data with {\sc niextract-events}. If two \nicer observations occurred during the \nustar observation, the GTIs were combined with {\sc ftmgtime}. The resulting event files were loaded into {\sc xselect} to extract \lc s in various energy bands. Source and background spectra were generated using the nibackgen3C50 tool \citep{remillard22} for each cleaned and ufa (calibrated but unfiltered) event file pair based on instrument proxies to account for the observing conditions at the time. Spectral response files were generated via {\sc nicerarf} and {\sc nicerrmf}.  

\begin{figure}[t!]
\begin{center}
\includegraphics[width=0.41\textwidth,trim=60 330 0 80,clip]{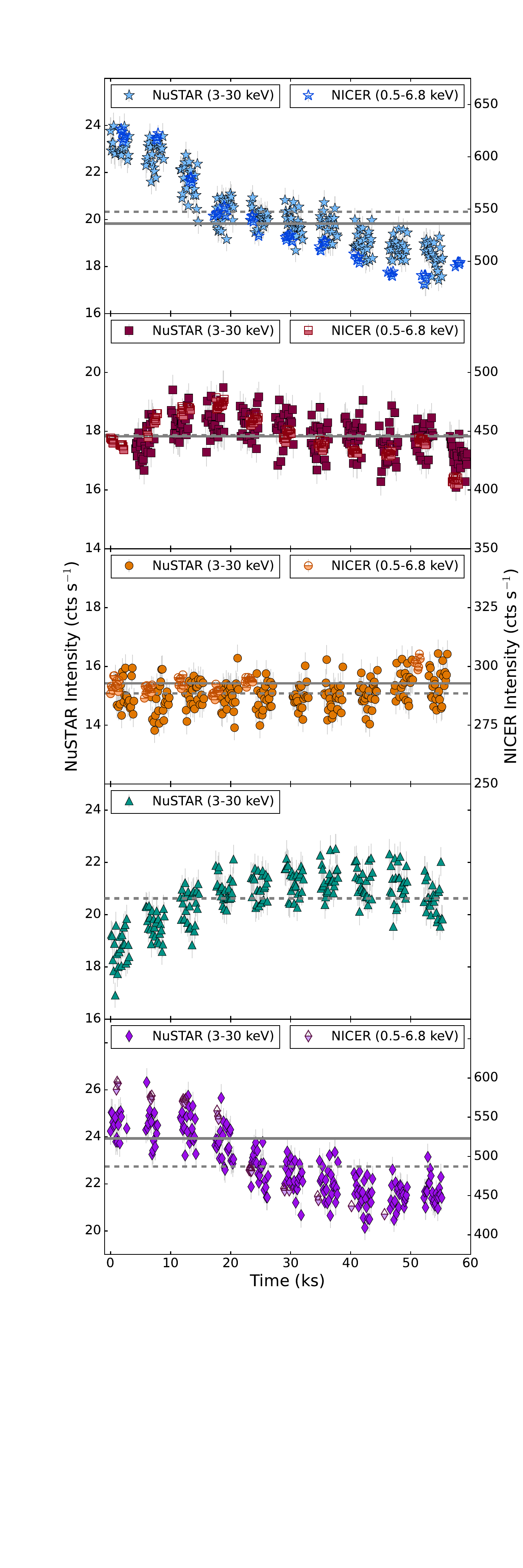}
\caption{Light curve for the \nustar/FPMA  and \nicer\ observations of \source binned to 128~s. Each panel represents a unique observation epoch, which are indicated by the unique symbols in each plot.The grey dashed and solid lines indicate the average count rate for \nustar\ and the \nicer, respectively. Only one FPM is shown for clarity.}
\label{fig:lc}
\end{center}
\end{figure}

\begin{figure}
\begin{center}
\includegraphics[width=0.48\textwidth,trim=5 0 0 0,clip]{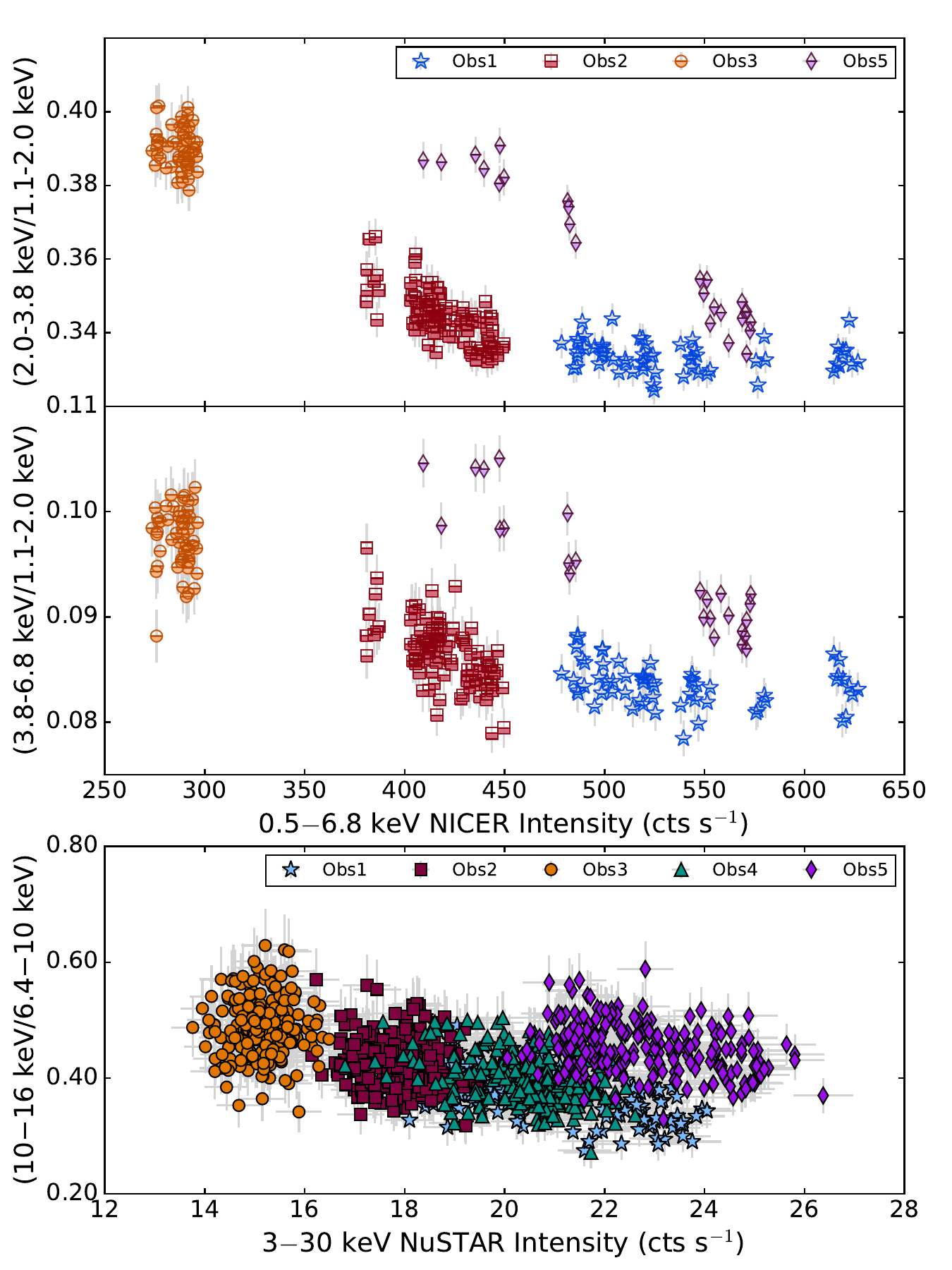}
\caption{The hardness-intensity diagrams for different  \nicer and \nustar energy bands for each joint observation of \source. The top two panels are constructed from \nicer data while the bottom panel is \nustar.}
\label{fig:HID}
\end{center}
\end{figure}

There were no Type-I X-ray bursts present in either data set, therefore no further filtering was needed. Systematic errors of 1\% in the full $0.3-10$ keV band were added to the \nicer spectrum. Both the \nicer and the \nustar spectra were binned using the optimal binning scheme \citep{kaastra16} with a requirement of at least 30 counts per bin to ensure the use of $\chi^{2}$ statistics. Figure~\ref{fig:lc} shows the \nustar and \nicer \lc s binned to 128~s for each observation. We show in Figure~\ref{fig:HID} the hardness ratio in two energy bands for \nicer\ ($2.0-3.8$~keV~/~$1.1-2.0$~keV and ($3.8-6.8$~keV~/~$1.1-2.0$~keV) versus the $0.5-6.8$ keV intensity of \source, as well as the hardness ratio ($10-16$~keV~/~$6.4-10$~keV) versus intensity from the \nustar band.

\section{Spectral Modeling and Results} \label{sec:methods}
In this section we describe the modeling techniques we use. We begin by modeling just the continuum, then we include \textsc{diskline} model components to attempt to account for the reflection features. After that we replace the \textsc{diskline} components with \xillverco and then we refine the use of \xillverco by tying certain parameters and fitting the data simultaneously.

\begin{table*}[h!t]
\begin{center}
\caption{Continuum Model, Individual Fit}
\label{tab:contind}

\begin{tabular}{lllllll}

\hline\hline
Model Component & Parameter                  & Obs1 & Obs2 & Obs3 & Obs4 & Obs5  \\  \hline \noalign{\vskip 1mm}    
CRABCOR         & C$_{FPMB}$ (10$^{-1}$)               &$10.00\pm0.04$   &$9.97\pm0.04$   & $9.89_{-0.05}^{+0.04}$    &$10.02\pm0.04$                         & $9.82\pm0.04$    \vspace{3mm}\\
                & C$_{NICER}$ (10$^{-1}$)              &9.3$^{+0.2}_{-0.1}$   &$9.7\pm0.2$   & $9.1_{-0.2}^{+0.3}$    & ---                         & $9.1_{-0.1}^{+0.3}$   \vspace{3mm}\\ 
                & $\Delta\Gamma$ (10$^{-2}$) &$-3.8_{-1.8}^{+1.2}$  &$1.0_{-1.6}^{+1.3}$   & $-2.5_{-1.1}^{+2.5}$   & ---                         & $-4.2_{-1.0}^{+2.3}$  \vspace{4mm} \\
EDGE            & E (10$^{-1}$ keV)          &$4.42\pm0.01$   &$4.33\pm0.01$   &$4.54\pm0.01$    & ---                         &                    $4.44\pm0.01$    \vspace{3mm}\\
                & $\tau_{max}$               &$2.8\pm0.1$   &$2.3\pm0.1$   &$2.3_{-0.1}^{+0.2}$    & ---                         & $3.5\pm0.1$   \vspace{4mm}\\
EDGE            & E (10$^{-1}$ keV)        &$8.62\pm0.01$   &$8.63\pm0.01$   &$8.49\pm0.02$             & ---                         &                    $8.63_{-0.02}^{+0.01}$    \vspace{3mm}\\
                & $\tau_{max}$ (10$^{-1}$)                &$4.4\pm0.1$   &$4.30_{-0.04}^{+0.05}$   &$3.0\pm0.1$    & ---                         & $4.1\pm0.1$   \vspace{4mm}\\
TBABS           &$N_{\rm H}$ (10$^{21}$ cm$^{-2}$)  &$2.3\pm0.1$   &$0.17\pm0.01$   & $0.07\pm0.02$    & ---                         &                    $0.011_{-0.001}^{+0.004}$  \vspace{3mm}\\
                
BBODY           & kT (keV)                   &$1.24\pm0.01$  &$1.20\pm0.02$   & $1.24\pm0.04$    &$1.20_{-0.01}^{+0.02}$                        & $1.13_{-0.01}^{+0.02}$   \vspace{3mm}\\
                & norm$_{bb}$ (10$^{-3}$)    &$2.5\pm0.1$   &$1.1\pm0.1$   &$0.8\pm0.1$    & $1.8\pm0.1$   & $1.7\pm0.1$   \vspace{4mm}\\
DISKBB          & kT (keV)                   &$0.427_{-0.007}^{+0.004}$   &$0.32\pm0.01$   &$0.64_{-0.01}^{+0.02}$    & ---                         &                     $0.45\pm0.01$   \vspace{3mm}\\
                & norm$_{dbb}$               & $731_{-38}^{+61}$        &$1088_{-140}^{+159}$        &$74_{-11}^{+6}$              & ---                         &$421^{+30}_{-60}$   \vspace{4mm}\\
POWERLAW        & $\Gamma$                   &$2.31_{-0.01}^{+0.02}$   &$2.29\pm0.01$   &$1.94\pm0.02$    &$2.30\pm0.01$   &                    $2.10\pm0.01$   \vspace{3mm}\\
                & norm$_{pl}$ (10$^{-1}$)                &$3.2_{-0.1}^{+0.2}$   & $3.3\pm0.1$   & $1.4\pm0.1$    & $3.6\pm0.1$   & $2.8\pm0.1$ \vspace{4mm}\\ \hline \noalign{\vskip 1mm}    
                &$\chi^2$ (dof)               &2410 (407)              &3705 (410)                   &954 (405)                     &374 (273)                     & 1264 (406)\\

\end{tabular}
\end{center}
\medskip
Note -- Errors are given at the 90\% level. The BBODY normalization is defined as $(L/10^{39}$ erg s$^{-1})/(D/10$ kpc)$^2$, the DISKBB normalization is defined as $(R_{in}/$km)$^2/(D/10$ kpc)$^2\times\cos{\theta}$, and the POWERLAW normalization is defined as photons keV$^{-1}$ cm$^{-2}$ s$^{-1}$ at 1 keV.
\end{table*}

\subsection{Continuum modeling} \label{subsec:cont}
We initially model the continuum with an absorbed three component model {\sc crabcor*edge*edge*tbabs*(bbody + diskbb + powerlaw)}. {\sc crabcor} is comprised of a multiplicative constant and a term $E^{-\Delta\Gamma}$ as done by \cite{steiner10}. This component aligns the amplitudes and slopes between \nicer and \nustar data in order to account for the mission specific calibration differences. The constant is held at 1 for the FPMA spectrum in \nustar. The $\Delta\Gamma$ component is fixed to zero for both FPMA and FPMB \nustar spectra, and allowed to vary in \nicer to align the slopes. The two edge components account for absorption edge features at low energies-- one at $\sim0.42$ keV and the other at  $\sim0.86$ keV -- similar to those used in \cite{ludlam20}. These absorption edges are likely astrophysical in origin, but we defer to previous studies which use \textsc{edge} in their analyses. {\sc tbabs} is a model component which accounts for absorption by the ISM along line of sight. For \textsc{tbabs}, we use the \textsc{wilm} abundance (\citealt{wilms00}) and the \textsc{vern} cross section (\citealt{verner96}). The three additive components {\sc bbody + diskbb + powerlaw} account for thermal emission from the NS, thermal emission from the disk, and non-thermal emission from the corona, respectively.

\begin{figure}[t!]
\begin{center}
\includegraphics[width=0.42\textwidth,trim=60 8 0 10,clip]{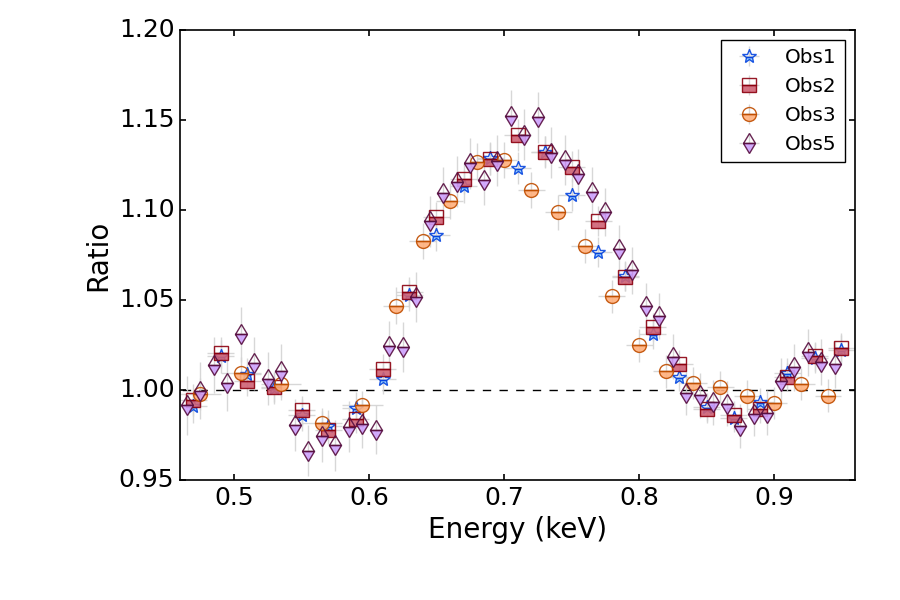}
\includegraphics[width=0.42\textwidth,trim=60 8 0 10,clip]{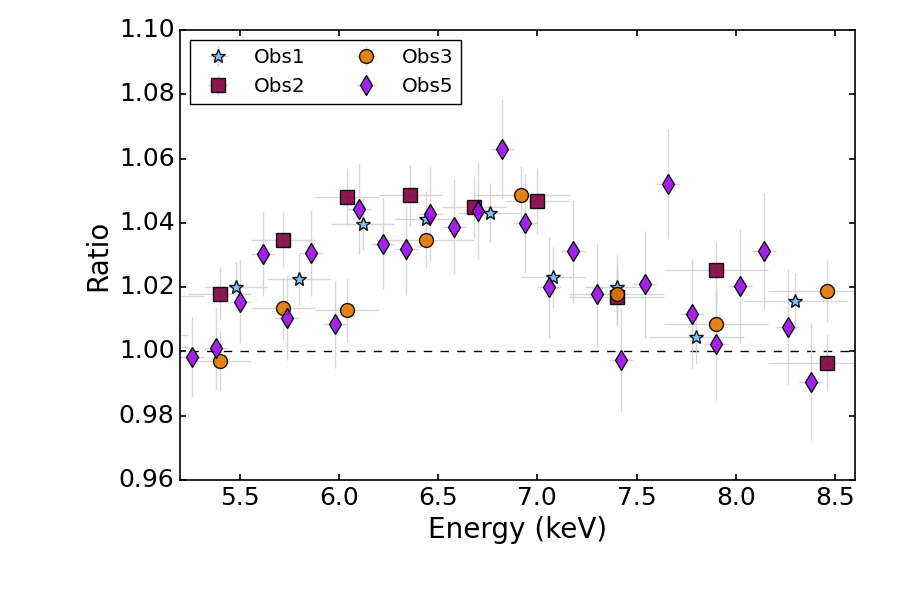}
\caption{(top) \ion{O}{8} Ly$\alpha$ line profile from \nicer for observations 1,2,3, and 5 centered at $\sim 0.7$ keV. (bottom) Fe K$\alpha$ profile from FPMA in \nustar centered at $\sim6.5$ keV. The Fe features are notably weaker than the \ion{O}{8} features, with the former peaking at around the 6\% above continuum, and the latter peaking around 15\% above continuum.}
\label{fig:lines}
\end{center}
\end{figure}

\begin{figure}[h!t]
\begin{center}
\includegraphics[width=0.39\textwidth,trim=55 90 20 60 ,clip]{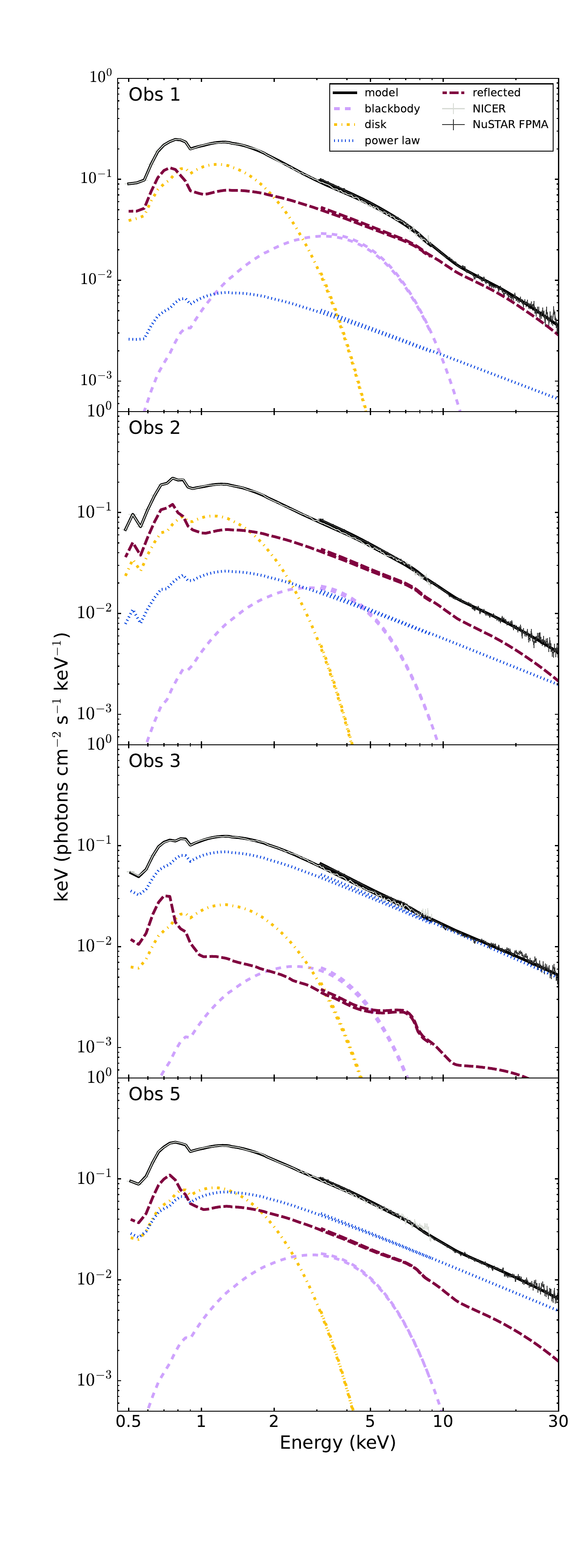}
\caption{The unfolded spectra and model components for each observation as the flux varies. The model components shown here are those displayed in Table \ref{tab:xillvercomb}. The power-law, displayed with a blue dotted line, becomes the dominant contributor to the overall spectrum  during Obs3, the lowest flux state.}
\label{fig:spectra}
\end{center}
\end{figure}

With the model in place, we use {\sc xspec} v.12.12.1 \citep{arnaud96} to fit each observation with the continuum. The initial fit is then used as a starting position for a Markov Chain Monte-Carlo (MCMC) fit. The chains are comprised of 100 walkers, with a burn-in period of 500,000 steps before a chain of 25,000 steps. The results of this fit is displayed in Table \ref{tab:contind}. Although the data are binned to a minimum number of counts per bin to allow the use of $\chi^2$ statistics, the  ``standard" weighting scheme in {\sc xspec} often resulted in over fitting the data, yielding a reduced $\chi^2<1$. This is a known issue discussed in \cite{galloway20} regarding the handling of the low-count bins in {\sc xspec} v.12. We therefore switch to the ``Churazov" weighting scheme when performing our spectral fits and report the values for the $\chi^2$/degree of freedom (dof) in Tables~2-5. Churazov weighting adjusts the weight by averaging the counts in surrounding channels, which can smooth the overall weighting and not apply disproportionate weight to local extrema \citep{churazov96}.  All errors are reported at the 90\% level. 

As can be seen in Table \ref{tab:contind}, observation 4 used a simplified version of the model. Since no \nicer data is available, {\sc crabcor} is replaced with a simple {\sc const} term to align the FPMA and FPMB of \nustar. In initial runs, attempting to use the full model resulted in poor constraints on parameters. This is due to the fact that much of the model information is primarily seen in the \nicer region of the spectrum. This includes both edge components and most of the absorption covered by {\sc tbabs}. For example, the {\sc diskbb} peak is always found to be $<0.7$ keV. An early fit using {\sc diskbb} had values for $kT$ = 0.35$^{+0.34}_{-0.04}$ keV and the normalization component was completely unconstrained, without any improvement in the $\chi^2$/dof. Observation 4 is well modeled by a very simple continuum, with a $\chi^2$/dof~$\approx 1.14$. As a result of this, we will not consider it for most of the remaining analysis.

Our continuum model does not provide a very good fit to observations 1--3 and 5, as is apparent from the reduced $\chi^2$ values in Table \ref{tab:contind}, and strong residuals are seen in the Fe~K and \ion{O}{8} regions. The shape of the line profiles in the residuals can be seen in Figure \ref{fig:lines}. These are extracted by first ignoring the regions surrounding the \ion{O}{8} ($0.60-0.80$~keV) and Fe~K ($5.5-7.4$~keV) features to avoid biasing the continuum fit. The continuum is then fit. The ignored regions are then reapplied to the spectrum to demonstrate that the continuum model is insufficient to encompass the reprocessed emission and another component is necessary to fully describe the X-ray spectrum.

\subsection{Modeling the Reprocessed Emission} \label{subsec:refl}

As an initial test of the strength of the reflected line features, we include two \textsc{diskline} components, centered near $0.67$ keV and $6.5$ keV. \textsc{diskline} is a model component which accounts for a single line emission feature from a relativistically blurred disk. The equivalent width EW of each line is measured using the \textsc{eqwidth} command in \textsc{xspec}. In addition to estimating the equivalent width, we also test the location of these features within the disk since the \ion{O}{8} Ly$\alpha$ and Fe K$\alpha$ features could arise from separate radii. However, due to the lower signal-to-noise of the Fe K$\alpha$ line,  the location of that feature is poorly constrained.  The lower limit on the emission radius is 6 gravitational radii (\rg~$=GM/c^2$) in each case, consistent with the \ion{O}{8} feature. We therefore tie the values of the emission radius \rin for both line components in the fit, the results of which can be found in Table \ref{tab:disklineind}.

To model the reprocessed emission we utilize {\sc relconv*xillverCO}. {\sc xillverCO} is a version of {\xillver}, an additive component for {\sc xspec} which is a table of synthetic reflected spectra \citep{garcia10, garcia13}, that has been developed with the unique features of UCXBs in mind, by including an overabundance of the carbon and oxygen expected from a white dwarf companion \citep{madej14}. {\sc relconv} is a relativistic blurring kernel used to account for the strong gravity close to the NS and Doppler broadening effects.

\begin{table*}[!t]
\begin{center}
\caption{Diskline Model, Individual}
\label{tab:disklineind}

\begin{tabular}{llllll}
\\ \hline\hline
Model Component & Parameter                 & Obs1 & Obs2 & Obs3 & Obs5 \\ \hline\noalign{\vskip 1mm}  
CRABCOR         & C$_{FPMB}$ (10$^{-1}$)                &$9.99_{-0.03}^{+0.05}$    &$9.98\pm0.04$    &$9.85_{-0.02}^{+0.08}$   & $9.84_{-0.04}^{+0.03}$   \vspace{3mm}\\
                & C$_{NICER}$ (10$^{-1}$)              &$8.5_{-0.2}^{+0.1}$    &$8.9_{-0.3}^{+0.2}$  & $9.0_{-0.3}^{+0.2}$     &$9.0_{-0.3}^{+0.2}$      \vspace{3mm}\\
                & $\Delta\Gamma$ (10$^{-2}$)  &$-7.5_{-1.5}^{+1.1}$   &$-4.7_{-1.9}^{+1.3}$    &$-2.6_{-2.2}^{+1.5}$    &$-5.6_{-2.0}^{+1.9}$     \vspace{4mm}\\
EDGE            & E (10$^{-1}$ keV)         &$3.5_{-0.5}^{+0.6}$    &$4.1_{-0.3}^{+0.2}$    &$3.3_{-0.1}^{+0.3}$     &$4.61_{-1.43}^{+0.03}$                      \vspace{3mm}\\
                & $\tau_{max}$              &$0.6_{-0.2}^{+0.3}$   &$0.9_{-0.1}^{+0.2}$    & $0.6\pm0.1$    &$1.0_{-0.1}^{+1.1}$      \vspace{4mm}\\
EDGE            & E (10$^{-1}$ keV)     &$8.98\pm0.03$   &$8.97_{-0.02}^{+0.03}$     &$8.79_{-0.04}^{+0.07}$     &$8.93_{-0.02}^{+0.06}$                      \vspace{3mm}\\
                & $\tau_{max}$ (10$^{-1}$)            & $1.93_{-0.05}^{+0.17}$   &$2.1\pm0.1$    &$1.32_{-0.03}^{+0.08}$     &$2.3_{-0.2}^{+0.1}$    \vspace{4mm}\\
TBABS           &$N_{\rm H}$(10$^{21}$ cm$^{-2}$) &$3.3\pm0.1$    &$2.63_{-0.04}^{+0.12}$    &$2.6\pm0.1$     &$2.3_{-0.1}^{+0.5}$                      \vspace{3mm} \\
                
BBODY           & kT (keV)                  &$1.21_{-0.02}^{+0.01}$    &$1.07\pm0.02$     &$1.04_{-0.04}^{+0.03}$     &$1.05_{-0.04}^{+0.01}$                      \vspace{3mm}\\
                & norm$_{bb}$ (10$^{-3}$)   &$2.52_{-0.19}^{+0.03}$    &$1.58_{-0.16}^{+0.03}$     &$0.8\pm0.1$     &$1.5\pm0.1$     \vspace{4mm}\\
DISKBB          & kT (keV)                  &$0.426_{-0.008}^{+0.004}$    &$0.408_{-0.008}^{+0.004}$     & $0.59\pm0.01$    &$0.417_{-0.024}^{+0.003}$                     \vspace{3mm}\\
                & norm$_{dbb}$              &$950_{-39}^{+74}$ &$730_{-60}^{+59}$ &$96_{-12}^{+16}$   &$639_{-4}^{+176}$     \vspace{4mm}\\
POWERLAW        & $\Gamma$                  &$2.29_{-0.01}^{+0.02}$    &$2.189_{-0.003}^{+0.027}$     &$1.96\pm0.02$     &$2.12_{-0.01}^{+0.02}$                      \vspace{3mm}\\
                & norm$_{pl}$ (10$^{-1}$)              &$3.1_{-0.1}^{+0.2}$    &$2.52_{-0.02}^{+0.19}$    &$1.5\pm0.1$     & $2.9_{-0.1}^{+0.2}$     \vspace{4mm}\\ 

DISKLINE$_1$    &E$_{line}$ (keV)           &$0.67\pm0.01$    &$0.692_{-0.004}^{+0.001}$     &$0.686_{-0.005}^{+0.004}$     & $0.649_{-0.007}^{+0.003}$  \vspace{3mm}\\
                & $\vert q\vert$             & $2.5\pm0.1$    &$2.36_{-0.02}^{+0.06}$     &$2.22_{-0.03}^{+0.10}$    &$2.4_{-0.2}^{+0.4}$    \vspace{3mm}\\
                & R$_{\rm in}$ ($R_{g}$)        &$6.4_{-0.4}^{+0.8}$    &$6.05_{-0.05}^{+0.29}$    &$6.3_{-0.3}^{+2.6}$      & $6.2_{-0.2}^{+0.1}$      \vspace{3mm}\\
                &norm (10$^{-2}$ keV)       &$7.3_{-0.7}^{+0.6}$    &$4.2_{-0.1}^{+0.4}$    & $1.5\pm0.1$      &$4.8_{-0.2}^{+0.6}$  \\
                &EW (eV)        &$63_{-7}^{+1}$ &$49_{-1}^{+4}$ &$34_{-3}^{+2}$ &$47_{-5}^{+2}$   \vspace{4mm}\\
DISKLINE$_2$    &E$_{line}$ (keV)           &$6.44_{-0.21}^{+0.08}$    &$6.59_{-0.16}^{+0.07}$     &$6.7\pm0.1$     &$6.3_{-0.2}^{+0.1}$  \vspace{3mm}\\
                &norm (10$^{-4}$ keV)        &$3.8_{-0.4}^{+1.5}$    &$5.0\pm1.0$    & $3.5_{-0.4}^{+0.5}$    &$7.0\pm1.0$\\
                &EW (eV)            &$70_{-20}^{+10}$   &$92_{-3}^{+25}$    &$90_{-10}^{+20}$ &$120_{-10}^{+20}$ \vspace{4mm}\\
                \hline\noalign{\vskip 1mm}
                &$\chi^2$ (dof)             &862 (401)                    &862 (404)                    &481 (399)                  &665 (400)
\end{tabular}
\end{center}
\medskip
Note -- Errors are given at the 90\% level. The BBODY normalization is defined as $(L/10^{39}$ erg s$^{-1})/(D/10$ kpc)$^2$, the DISKBB normalization is defined as $(R_{in}/$km)$^2/(D/10$ kpc)$^2\times\cos{\theta}$. The POWERLAW normalization is defined as photons keV$^{-1}$ cm$^{-2}$ s$^{-1}$ at 1 keV and the DISKLINE normalization is defined as photons cm$^{-2}$ s$^{-1}$.
\end{table*}

In {\sc relconv}, the two emissivity indices $q_1$ and $q_2$ for the inner and outer disk are tied and reported as $q$, and so the radius $R_{br}$ at which these indices differ becomes obsolete. The outer radius is fixed at 990~$R_g$, and the inner disk radius \rin in units of $R_{ISCO}$ ($1~R_{ISCO} = 6~R_g$ for spin $a=0$) is free to vary. The limb, spin, and redshift parameters are all fixed at zero. In {\sc xillverCO}, we tie the value of $\Gamma$ to the power-law index of the continuum power-law component as this is the illuminating input source of the reprocessed emission in the model. The carbon and oxygen abundance $A_{CO}$ is measured in units of solar abundance, and the value denoted as Frac is the measured ratio of the illuminating power-law flux to the flux of the emergent blackbody. The normalization of \xillverco is defined such that an incident spectrum with flux $F(E)$ follows the expression 
\begin{equation}
    \int_{0.1\,\mathrm{keV}}^{1 \,\mathrm{MeV}}F(E)dE = \frac{10^{35}}{4\pi}\,\mathrm{erg}\,\mathrm{cm}^4\,\mathrm{s}^{-1}
\end{equation}
(\citealt{dauser16}). The inclination of the system is fixed at 55$^{\circ}$ because of the high number of free variables. This value is consistent with the inclinations measured by \cite{madej14} and \cite{ludlam19a}. We test that our results are not dependent upon this choice of fixed inclination by running fits with the inclination tied across observations but free to vary. The inclination  remains consistent with  55$^{\circ}$. We find that all key parameters of interest are consistent within the 90\% confidence level to the fits with fixed inclination. Therefore, our results are robust with respect to inclination and we only report on the fits with this parameter fixed.

Initially the model is fit to each observation, using the same methods as discussed in Section \ref{subsec:cont}. 
Once fit, the ionization parameter $\xi$ of the system is calculated with the value of Frac and the disk temperature $kT$ from {\sc xillverCO} using: 
\begin{equation}
    \xi = \frac{4\pi}{n} F_x
\end{equation}
where $F_x = {\rm Frac} \times \sigma T^4$, defined using the blackbody temperature and Frac as defined in \xillverco. Note that the blackbody temperature from \xillverco differs from the disk temperature in \textsc{diskbb} in that it represents the region of the disk where reflection occurs versus the emission of the entire disk as measured by \textsc{diskbb}. \textsc{diskbb} is a multi-temperature blackbody representing emission from the entire disk, whereas the thermal temperature included in \xillverco only accounts for disk emission at the radius at which the O line originates\footnote{See \cite{madej14} for further discussion.}. The final value is reported in log units. Here, $n$ is the disk number density that is a fixed value of $1\times10^{17}$ cm$^{-3}$ in the reflection model. The results of these fits and calculations are reported in Table~\ref{tab:xillverind}.

\begin{table*}[h!t]
\begin{center}
\caption{Reflection Model Fits, Individual}
\label{tab:xillverind}

\begin{tabular}{llllll}
 \hline\hline
Model Component & Parameter                 & Obs1 & Obs2 & Obs3 & Obs5 \\ \hline\noalign{\vskip 1mm}  
CRABCOR         & C$_{FPMB}$(10$^{-1}$)                &$9.99_{-0.04}^{+0.05}$            &$9.99_{-0.06}^{+0.02}$     &$9.88_{-0.03}^{+0.06}$     &$9.84_{-0.05}^{+0.03}$                          \vspace{3mm}\\
                & C$_{NICER}$(10$^{-1}$)               &$8.8_{-0.1}^{+0.4}$    &$8.6_{-0.1}^{+0.3}$  &$8.9_{-0.2}^{+0.3}$     &$8.6_{-0.3}^{+0.2}$      \vspace{3mm}\\
                & $\Delta\Gamma$ (10$^{-2}$)  & $-5.2_{-0.7}^{+2.6}$   &$-7.2_{-1.0}^{+2.0}$    &$-3.5_{-1.6}^{+2.2}$    &$-8.3_{-1.8}^{+1.5}$      \vspace{4mm}\\
EDGE            & E (10$^{-1}$ keV)        & $4.2_{-0.9}^{+0.2}$    &$4.2_{-0.4}^{+0.2}$     &$3.6_{-0.5}^{+1.0}$     & $4.20_{-0.7}^{+0.4}$                      \vspace{3mm}\\
                & $\tau_{max}$              &$0.4_{-0.1}^{+0.4}$    &$0.8\pm0.1$     & $0.5\pm0.2$    &$0.50_{-0.08}^{+0.44}$      \vspace{4mm}\\
EDGE            & E (10$^{-1}$ keV)         &$8.60_{-0.03}^{+0.05}$    &$8.68_{-0.02}^{+0.03}$     &$8.55_{-0.03}^{+0.08}$    &$8.69_{-0.05}^{+0.02}$                      \vspace{3mm}\\
                & $\tau_{max}$ (10$^{-1}$)                &$2.0_{-0.2}^{+0.1}$    &$2.9\pm0.1$     & $1.3_{-0.1}^{+0.2}$     &$2.6_{-0.1}^{+0.2}$      \vspace{4mm}\\
TBABS           & $N_{\rm H}$ (10$^{21}$ cm$^{-2}$) &$3.5_{-0.1}^{+0.2}$    &$3.4\pm0.1$    &$3.3_{-0.2}^{+0.1}$     &$3.5_{-0.2}^{+0.1}$                      \vspace{3mm} \\
                
BBODY           & kT (keV)                  &$1.09\pm0.02$    &$1.13\pm0.02$     &$0.77_{-0.06}^{+0.04}$     &$1.09_{-0.05}^{+0.01}$                      \vspace{3mm}\\
                & norm$_{bb}$ (10$^{-3}$)   &$2.1_{-0.1}^{+0.2}$    &$1.1\pm0.1$     &$0.7\pm0.1$     & $0.9\pm0.1$     \vspace{4mm}\\
DISKBB          & kT (keV)                  &$0.40\pm0.01$    &$0.36\pm0.01$     &$0.50_{-0.03}^{+0.02}$     &$0.37\pm0.01$                     \vspace{3mm}\\
                & norm$_{dbb}$              &$1545_{-92}^{+98}$          &$926\pm67$             &$131_{-27}^{+39}$              &  $832_{-88}^{+144}$     \vspace{4mm}\\
POWERLAW        & $\Gamma$                  &$2.07_{-0.06}^{+0.03}$    &$2.33_{-0.02}^{+0.01}$     &$2.05\pm0.01$     &$2.25_{-0.02}^{+0.01}$                       \vspace{3mm}\\
                & norm$_{pl}$ (10$^{-1}$)               &$0.6_{-0.3}^{+0.1}$    &$3.0\pm0.1$     & $1.6\pm0.1$     & $3.6_{-0.2}^{+0.1}$     \vspace{4mm}\\ 

RELCONV         & q                         &$2.5_{-0.1}^{+0.2}$    &$2.6\pm0.6$     &$2.4\pm0.2$     &$2.6\pm0.1$                   \vspace{3mm}\\
                & \rin (\risco)     &$1.04_{-0.04}^{+0.32}$    &$1.03_{-0.03}^{+0.13}$     &$1.5_{-0.5}^{+0.3}$     &$1.03_{-0.03}^{+0.16}$      \vspace{4mm}\\
XILLVER$_{CO}$  & A$_{CO}$                  &$36.0_{-2.9}^{+3.9}$   &$19.5_{-2.4}^{+2.5}$    &$11.0_{-1.9}^{+3.9}$     &$21.1_{-5.9}^{+0.3}$                     \vspace{3mm}\\
                & kT$_{bb}$ (10$^{-2}$ keV)     &$9.9_{-0.4}^{+0.1}$    &$5.67_{-0.07}^{+0.43}$     &$5.8_{-0.2}^{+0.3}$     &$5.45_{-0.02}^{+0.32}$     \vspace{3mm}\\
                & Frac (10$^{-1})$                     &$2.4_{-0.2}^{+0.4}$    &$0.14_{-0.02}^{+0.01}$    &$0.4\pm0.1$     &$0.16_{-0.03}^{+0.02}$      \vspace{3mm}\\
                & norm (10$^{-9}$)          &$2.0_{-0.2}^{+0.3}$    &$209.2_{-53.8}^{+14.4}$ &$27.3_{-5.5}^{+6.1}$    &$250.7_{-64.0}^{+27.7}$     \vspace{3mm}\\ 
                &$\log(\xi)$(erg cm/s)      &$3.5\pm0.1$              &$1.3\pm0.1$              &$1.7\pm0.1$               &$1.3\pm0.1$      \vspace{4mm}\\
                \hline\noalign{\vskip 1mm} 
                
                &$\chi^2$ (dof)             &647 (401)                   &796 (404)                     &442 (399)                     &461 (420)
\end{tabular}
\end{center}
\medskip
Note -- Errors are given at the 90\% level. The BBODY normalization is defined as $(L/10^{39}$ erg s$^{-1})/(D/10$ kpc)$^2$, the DISKBB normalization is defined as $(R_{in}/$km)$^2/(D/10$ kpc)$^2\times\cos{\theta}$, and the POWERLAW normalization is defined as photons keV$^{-1}$ cm$^{-2}$ s$^{-1}$ at 1 keV.
\end{table*}

After fitting each observation individually with the aforementioned models, we simultaneously fit all observations tying components that should remain consistent over time.  These consistent variables are the hydrogen column density $N_H$, the absorption edge components, the slope difference between \nustar and \nicer $\Delta\Gamma$, and the carbon-oxygen abundance $A_{CO}$. With these parameters tied, we repeat the same process as before, and report the results in Table \ref{tab:xillvercomb}. Compared to the continuum model without accounting for relativistically broadened reflection features, the $\chi^2/$dof improves greatly.

We also measure the unabsorbed flux of each component between $0.5-50$~keV as the system varies. This range is chosen to be consistent with \cite{ludlam19a} to allow for direct comparison to other accreting sources. The trends for the blackbody, disk blackbody, and reflected emission follow the same trend of the overall flux. The power-law flux representing the coronal emission, however, seems to vary in the opposite direction, increasing in flux as the overall flux decreases, as shown in Figure \ref{fig:spectra}. These values are reported in Table \ref{tab:fluxes}.

\begin{table*}[h!t]
\begin{center}
\caption{Reflection Model, Combined}
\label{tab:xillvercomb}

\begin{tabular}{llllll}

\hline\hline
Model Component & Parameter                 & Obs1 & Obs2 & Obs3 & Obs5 \\ \hline\noalign{\vskip 1mm}  
CRABCOR         & C$_{FPMB}$ (10$^{-1}$)            &$10.00\pm0.01$     &$9.971_{-0.005}^{+0.004}$     &$9.90\pm0.02$     &$9.828_{-0.009}^{+0.003}$     \vspace{3mm}\\
                & C$_{NICER}$ (10$^{-1}$)               & $9.12\pm0.01$      &$9.134_{-0.003}^{+0.006}$      & $9.02\pm0.01$     &$9.29\pm0.01$      \vspace{3mm}\\
                & $^{\dagger}\Delta\Gamma$ (10$^{-2}$)  &\multicolumn{4}{c}{$-2.975_{-0.011}^{+0.004}$}     \vspace{4mm}\\
EDGE            & $^{\dagger}$E (10$^{-1}$ keV)                   &\multicolumn{4}{c}{$4.10_{-0.02}^{+0.05}$}   \vspace{3mm}\\
                & $^{\dagger}\tau_{max} $(10$^{-1}$)              &\multicolumn{4}{c}{$3.55_{-0.13}^{+0.02}$} \vspace{4mm}\\
EDGE            & $^{\dagger}$E (10$^{-1}$ keV)                   &\multicolumn{4}{c}{$8.720_{-0.004}^{+0.007}$}    \vspace{3mm}\\
                & $^{\dagger}\tau_{max}$ (10$^{-1}$)           &\multicolumn{4}{c}{$1.661_{-0.002}^{+0.008}$}  \vspace{4mm}\\
TBABS           & $N_{\rm H}$~$^{\dagger}$ (10$^{21}$ cm$^{-2}$)  &\multicolumn{4}{c}{$3.27\pm0.01$}       \vspace{4mm}\\
BBODY           & kT (10$^{-1}$ keV)                  &$10.76\pm0.01$    &$9.20\pm0.01$     &$8.01_{-0.05}^{+0.03}$     &$9.55_{-0.05}^{+0.01}$     \vspace{3mm}\\
                & norm$_{bb}$ (10$^{-3}$)   &$2.829_{-0.002}^{+0.012}$    &$1.59_{-0.02}^{+0.01}$     & $0.50\pm0.01$     &$1.640_{-0.003}^{+0.007}$      \vspace{4mm}\\
DISKBB          & kT (10$^{-1}$ keV)                  &$4.230\pm0.003$    & $3.803_{-0.009}^{+0.003}$     &$5.09_{-0.02}^{+0.03}$     &$3.93_{-0.02}^{+0.01}$     \vspace{3mm}\\
                & norm$_{dbb}$              &$1347_{-4}^{+8}$ &$1391_{-3}^{+11}$ &$113_{-4}^{+3}$ &$1071_{-8}^{+27}$     \vspace{4mm}\\
POWERLAW        & $\Gamma$                  &$1.912_{-0.003}^{+0.001}$     &$1.963_{-0.001}^{+0.003}$     &$2.059_{-0.001}^{+0.003}$     &$1.991_{-0.004}^{+0.002}$      \vspace{3mm}\\
                & norm$_{pl}$ ($10^{-2}$)               &$1.44_{-0.01}^{+0.03}$    &$5.1\pm0.1$     &$17.9_{-0.1}^{+0.2}$     &$14.2_{-0.3}^{+0.1}$      \vspace{4mm}\\ 

RELCONV          & q                         &$2.695_{-0.005}^{+0.010}$    &$2.672_{-0.004}^{+0.009}$     &$2.72_{-0.01}^{+0.02}$     &$2.728_{-0.003}^{+0.004}$      \vspace{3mm}\\
                & R$_{in}$ (R$_{ISCO}$)     & $1.01_{-0.01}^{+0.04}$    &$1.003_{-0.003}^{+0.038}$     &$1.92_{-0.04}^{+0.05}$     &$1.04\pm0.01$      \vspace{4mm}\\
XILLVER$_{CO}$  & $^{\dagger}$A$_{CO}$                  &\multicolumn{4}{c}{$25.70_{-0.07}^{+0.27}$ }     \vspace{3mm}\\
                & kT$_{bb}$ (10$^{-2}$ keV)     &$7.985_{-0.008}^{+0.050}$    & $6.22_{-0.03}^{+0.01}$     &$6.51_{-0.16}^{+0.09}$     &$5.89_{-0.02}^{+0.09}$      \vspace{3mm}\\
                & Frac (10$^{-1}$)                   & $5.07_{-0.11}^{+0.02}$    &$5.85_{-0.02}^{+0.10}$     &$0.49\pm0.01$     &$4.63_{-0.14}^{+0.07}$      \vspace{3mm}\\
                & norm (10$^{-9}$)          &$2.27_{-0.02}^{+0.01}$    &$3.46_{-0.04}^{+0.03}$     &$6.8_{-0.4}^{+0.7}$     &$4.4_{-0.2}^{+0.1}$ \vspace{3mm}\\     
                &$\log(\xi) (erg cm/s)$     &$3.42\pm0.01$              &$3.05\pm0.01$                &$2.05\pm0.02$              &$2.86\pm0.02$

                \vspace{4mm}\\ \hline\noalign{\vskip 1mm}  
                &$\chi^2$ (dof)             & \multicolumn{4}{c}{2593 (1625)}                    
\end{tabular}
\end{center}
\medskip
Note -- Errors are given at the 90\% level. The BBODY normalization is defined as $(L/10^{39}$ erg s$^{-1})/(D/10$ kpc)$^2$, the DISKBB normalization is defined as $(R_{in}/$km)$^2/(D/10$ kpc)$^2\times\cos{\theta}$, and the POWERLAW normalization is defined as photons keV$^{-1}$ cm$^{-2}$ s$^{-1}$ at 1 keV. $^{\dagger}$: Tied parameters. 
%\textbf{Some units vary by an order of magnitude when compared to Tables \ref{tab:xillverind} and \ref{tab:disklineind} for readability.}
\end{table*}

Across each observation, \rin is consistent with $1~ R_{ISCO}$. The exception is in the lowest flux observation, Obs3, for which the disk appears to recede to $1.92 R_{ISCO}$. To statistically test this result, we repeat the fit in Table \ref{tab:xillvercomb}, but this time we fix \rin to $1 R_{ISCO}$. The quality of the overall fit worsens by $3.9\sigma$ ($\Delta\chi^{2}=15$ worse for 1 dof), hence the disk truncation is required by the data. Over the course of the observations, the blackbody temperature and $\log{(\xi)}$ correlates with the overall flux, increasing as the flux of the system increases and vice versa. The temperature of the disk shows no such correlation, remaining relatively constant across the observations 1, 2, and 5, but increases significantly during our lowest flux observation. This contributes to the increased hardness during Obs3. 

We also recognize that the low disk normalization and high disk temperature of Obs 3 in Table \ref{tab:xillvercomb} implies a smaller inner disk radius (4.5 km assuming $\theta = 55^o$ and $D=3.3$kpc). Even when accounting for a spectral hardening factor, which can increase the calculated inner disk radius by a factor of $\sim1.6$ (\citealt{shimura95,lazar21}), we find the inferred inner disk radius to be far smaller than the truncated inner disk radius from \textsc{relconv}. This is possibly caused by a degeneracy between the disk and other components. We find that if we force the disk normalization to a value more consistent with the other observations, the disk temperature decreases to well below our pass band and the blackbody temperature increases to a value that is much too high (e.g., it increased to  roughly 9 keV in some cases). We do not assume this to be a physical parameter of the system at this observation. Additionally, we tested the use of {\sc simpl} \citep{steiner09} instead of {\sc powerlaw} to account for Comptonization in the continuum  while remaining consistent with the input model of \xillverco in an attempt to correct for the inconsistent behavior of the {\sc diskbb} component in Obs 3. This also led to a truncated disk in Obs 3, though the value was poorly constrained. The fit quality was greatly reduced (with the reduced $\chi^2$ nearly doubling); therefore we do not report on them further.

For completeness we also tried other model components before using those in the final table. For example, we attempted to replace \textsc{tbabs*edge*edge} with \textsc{varabs}, a model component with variable absorption that could account for both edges as well as any other non-solar abundance in the ISM. Tests with this component yielded results that were similar to those in the final analysis, but with the added complication of greatly increasing the number of free parameters (\textsc{tbvarabs} has up to 42 free parameters). Because of the already large number of free parameters in the tied fits shown in Table \ref{tab:xillvercomb}, we opted to use the simpler of the two models.

\begin{table*}[h!t]
\begin{center}

\caption{Component fluxes}
\label{tab:fluxes}

\begin{tabular}{lllll}

\hline\hline
Model Component          & Obs1 & Obs2 & Obs3  & Obs5  \\  \hline \noalign{\vskip 1mm}    
Blackbody                &$2.4\pm0.1$ &$1.4\pm0.1$ &$0.4\pm0.1$ &$1.3\pm0.1$ \\
Disk                     &$6.4\pm0.2$ &$4.1\pm0.1$ &$1.3\pm0.2$ &$3.6\pm0.2$ \\
Powerlaw                 &$1.3\pm0.3$ &$4.1\pm0.5$ &$12.1\pm0.3$ &$10.8\pm1.2$ \\
XillverCO                &$12.0\pm1.0$ &$9.8\pm1.2$ &$1.4\pm2.7$ &$7.6\pm1.8$ \\
Total                    &$19.5\pm1.1$ &$17.3\pm1.3$ &$13.7\pm2.7$ &$21.6\pm2.2$ \\

\end{tabular}
\end{center}
\medskip
Note.--- All fluxes reported have units $10^{-10}$ ergs s$^{-1}$ cm$^{-2}$ and are measured in the range $0.5-50$~keV. It should be noted that Obs5 occurred approximately 3 months after the first four observations.
\end{table*}

\section{Discussion} \label{sec:discussion}

We perform a spectral analysis of four simultaneous \nicer and \nustar observations of \source.
Each observation shows the presence of disk reflection in the form of emission line features that have been relativistically broadened indicating that they originate from the innermost disk region. 
We find that the disk shows evidence of truncation during Obs 3. We consistently find an increase in the inner disk radius for this observation compared to the other higher flux observations. In particular, when we simultaneously fit all observations, tying parameters that should remain physically unchanged between them, observation 3 is the only instance where \rin is inconsistent with 1 ~\risco (see Table~\ref{tab:xillvercomb}).

\begin{figure*}[h!t]
\begin{center}
\includegraphics[width=0.9\textwidth,trim=0 0 0 0,clip]{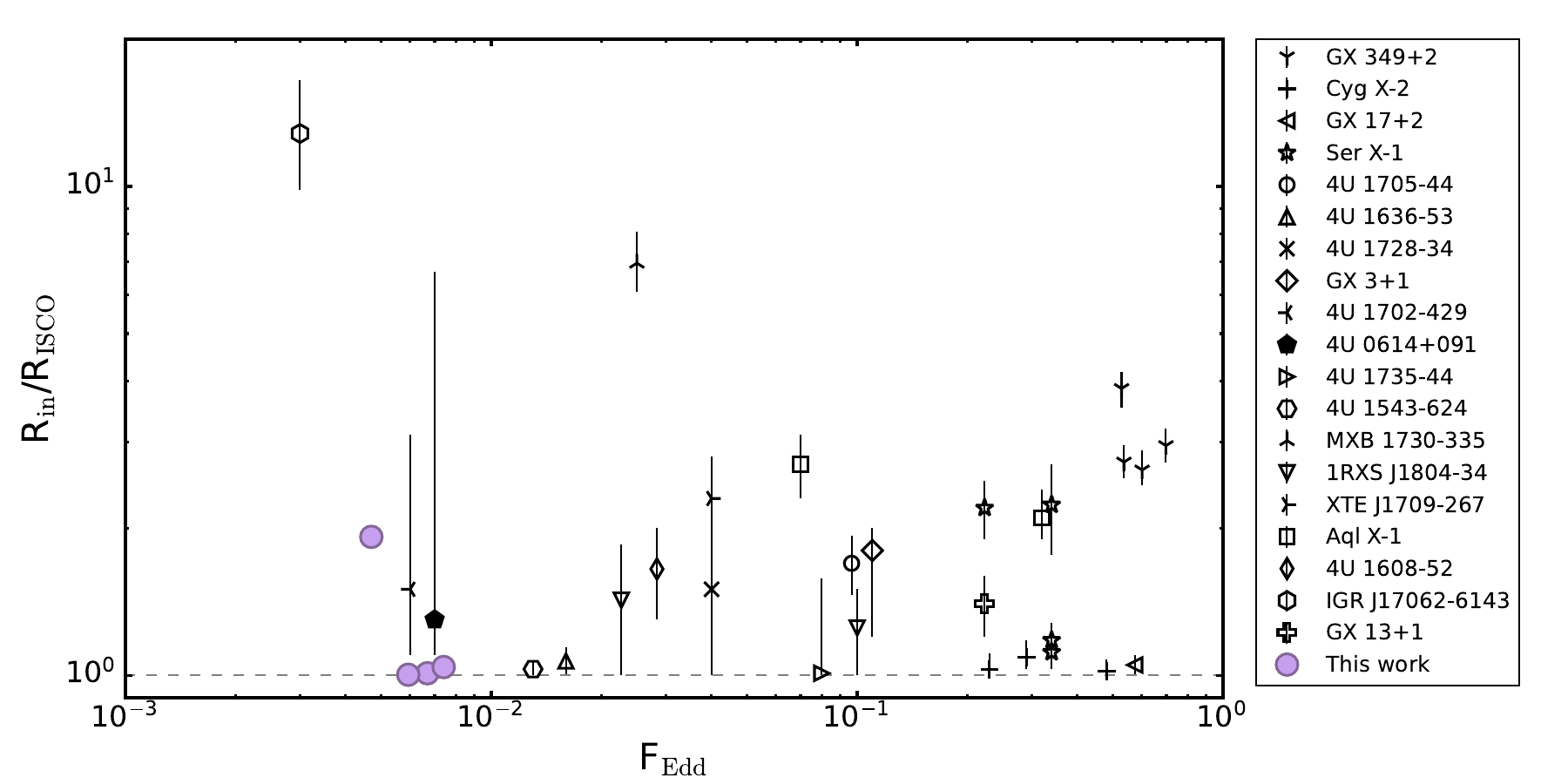}
\caption{Eddington fraction $F_{Edd}$ and the location of the inner disk as measured for several NS LMXBs studied with \nustar, and compare to this work. The Eddington fraction is the given luminosity of the source compared to the empirical Eddington luminosity for a NS. We see that there is almost no correlation between F$_{Edd}$ and the inner disk radius, which differs from the more rigid correlation} seen in BH LMXBs. A previous measurement of \rin from reflection modeling of NuSTAR data for this source is filled in black. 
\label{fig:feddrin}
\end{center}
\end{figure*}

Because we only observed one minimum during the long term light curve, it is difficult at this time to determine whether the apparent disk truncation is correlated with the low flux state, or if it is a coincidence. However, this behavior is analogous to that seen in BH LMXBs. In many BH LMXBs, it is expected that during the low/hard state, this type of truncation might be observed \citep{done07}. In these BH LMXB systems, the disk recedes at values of L/L$_{Edd}$ below $\sim0.01$ \citep{tomsick09}. By examining the contributing fluxes of each component as shown in Figure \ref{fig:spectra}, we can see that during the lowest flux state, the flux of the power-law component becomes dominant, while the thermal sources of X-rays (as well as the reflected emission) becomes a small fraction of the overall flux. Using the empirical Eddington luminosity for a 1.4 M$_{\odot}$ NS $L_{Edd} = 3.8\times10^{38}$~\lumcgs \citep{kuulkers03}, we compile archival measured values \citep{ludlam19a, ludlam20, mondal20, ludlam21, saavedra23} for \rin for various NS LMXBs studied with NuSTAR in Figure \ref{fig:feddrin} and compare to their Eddington fractions at time of observation. This figure demonstrates that, while there may be some correlation between the Eddington fraction F$_{Edd}$ and \rin, there is not a direct correlation for NSs, though similar behavior can be seen for the NS LMXB Ser X-1 \citep{chiang16, mondal20}. In that source, the authors report evidence of disk truncation during a low flux state. In this study, we are able to probe some of the lowest regimes of the Eddington ratio, and we find our result to be consistent with  results on the same source from \cite{ludlam19a} at the same flux as observation 1, with a major difference being the lack of \nicer data in the initial study. Since NSs and BHs are both compact accretors, we might expect their behavior to be similar. This is complicated by the magnetic fields and surface that are present around a NS. 

To test the feasibility of a disk depletion between observations 2 and 3, we calculate the mass accretion rate $\dot{M}$ using the equation:
\begin{equation}
    \dot{M} = \frac{L}{\eta c^2}
\end{equation}
where $L$ is the luminosity, calculated using the unabsorbed flux between 0.5-50 keV at 3.3 kpc (this distance was obtained by \citealt{arnason21} using GAIA DR2), and we assume an efficiency $\eta=0.2$. To obtain the minimum time needed to deplete the inner region between $1-2$~\risco, we use the minimum measured flux for observation 3 and hence the lowest observed mass accretion rate, allowing for a more conservative estimate. We calculate a value $\dot{M} = 4.37\times10^{-18} M_{\odot} {\rm yr}^{-1}$, and then combine with the Shakura-Sunyaev disk solution for the surface density $\Sigma$ to calculate the disk mass $M_{disk}$ (see \citealt{frank02} for more detail): 
\begin{equation}
    M_{disk} = 2\pi\int_{R_{*}}^{R_{out}}\Sigma R dR
\end{equation}

We find that at this conservative $\dot{M}$, disk truncation due to a complete depletion would take far longer than the $\sim2$ days between observations 2 and 3 (approximately 5 years). This implies that disk depletion is not the primary driver of truncation in this system. Instead it may be related to a change in the transition radius between the accretion disk and the coronal flow, as discussed for black hole systems in \cite{liu99}. This is supported by an increased flux of the coronal emission during this low flux state. In \cite{marino20}, the authors conduct a study of \source and suggest that during an ``extreme island state", atoll sources often have lower blackbody temperatures and truncated disks. This is consistent with what we are seeing in observation 3, where the blackbody temperature has decreased and the spectrum is harder than what is seen in the other observations. In both \cite{marino20} and \cite{migliari10}, the authors discuss the coupling between the jet and the disk. Unfortunately without simultaneous radio measurements, we can not know whether this disk behavior is driven by a jet.

The disk could be truncated due to the magnetic field strength of the NS at such a low mass accretion rate. To estimate the equatorial strength of the magnetic field, we can use the following equation, as done in \cite{ludlam20}:

\begin{multline}
    B = 3.5 \times 10^5\ k_A ^{-7/4} x^{7/4}\left( \frac{M}{1.4\ M_{\odot}} \right)^2 \left(\frac{10\ \rm km}{R_{\rm NS}}\right)^3 \\
    \times \left(\frac{f_{ang}}{\eta} \frac{F_{bol}}{10^{-9}\ {\rm erg s^{-1}cm^{-2}}} \right)^{1/2} \frac{D}{3.5\ \rm kpc}\ \rm G
\end{multline}
where $\eta$ is assumed to be 0.2, the conversion factor $k_A$ and the angular anisotropy $f_{ang}$ are set to unity, the distance $D$ to \source is 3.3 kpc (as mentioned previously), canonical NS values are used for $M$ and $R_{NS}$, and the $0.5-50$ keV flux and the inner disk radius (in $R_g$) for observation 3 are used for $F_{bol}$ and $x$ respectively. We obtain an upper limit on the magnetic field to be $B \leq 0.6\times 10^8$~G at the equator or $B \leq 1.2\times 10^8$~G at the pole. The magnetosphere radius, assumed to be proportional to $\dot{M}^{-2/7}$, does not follow the measured \rin, so the disk truncation is not driven primarily by the magnetosphere either, similar to what is seen in \cite{chiang16}.

\begin{figure}[!t]
\begin{center}
\includegraphics[width=0.46\textwidth,trim=67 27 10 10,clip]{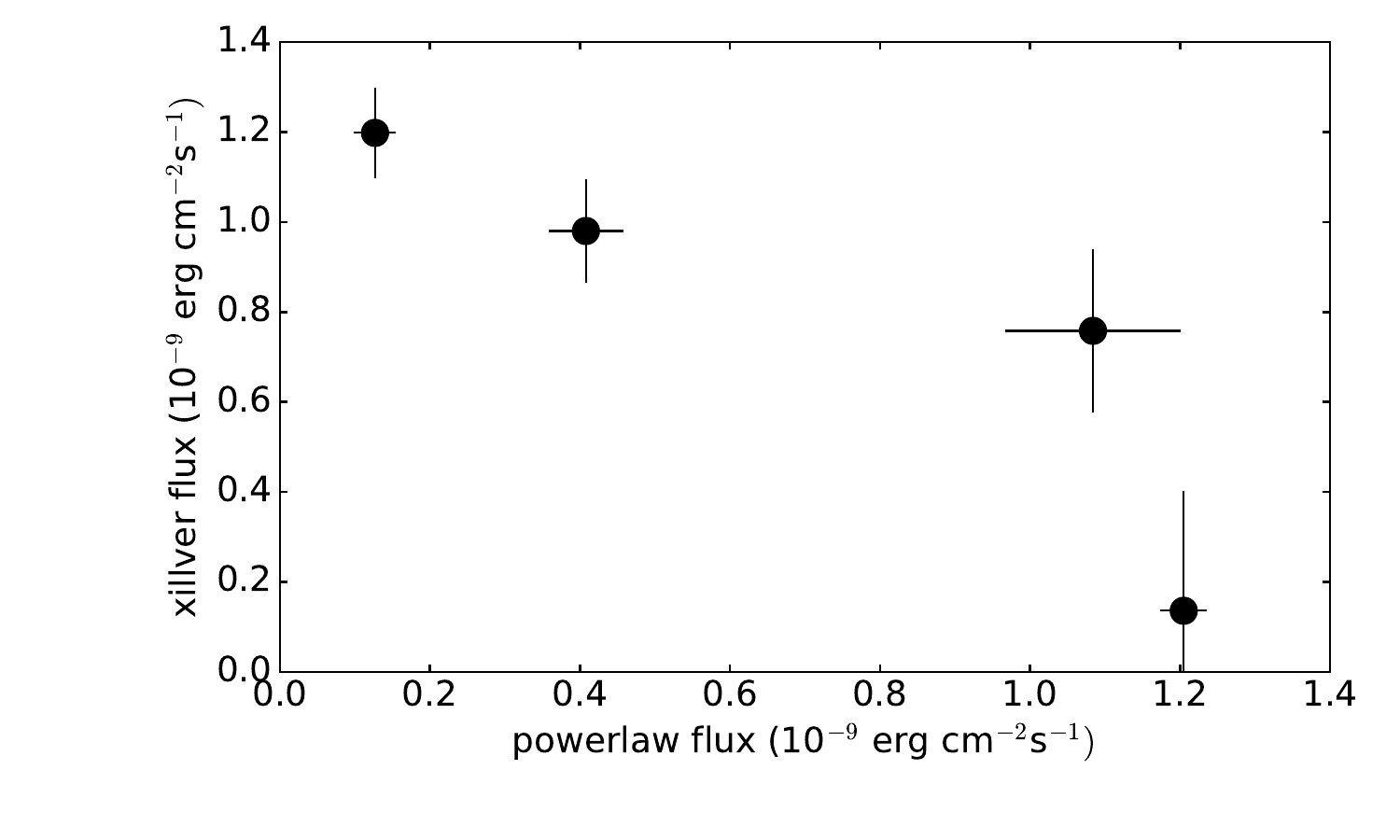}
\caption{The plot above demonstrates that while there appears to be some negative correlation between the flux of the power-law component and the flux of the reflected emission, the 90\% error bars show that across all observations except observation 3, which is the right-most point, the reflected emission is relatively consistent. }
\label{fig:plvsxillver}
\end{center}
\end{figure}

Because the model assumes that the illuminating flux is provided by the X-ray corona, the inverse relationship between the reflected flux and the power-law flux may seem contrary. Here we provide two potential explanations for this.
\begin{enumerate}
\item \textsc{xillverCO} is a stand-alone reflection table, hence it does not simultaneously model the input continuum component and emergent reprocessed emission. 
We are utilizing a simple power-law component in the continuum and tie the photon index of the reflected emission to it in order to account for the incident flux, then broadening the features of \textsc{xillverCO} using \textsc{relconv}. This could lead to potential degeneracy between the reflected emission and the incident emission or with the thermal components, leading to some skewing of the relative contributions of each component. However, we believe the overall trends hold true given that if the relative contributions were skewed due to some degeneracy between model components, then they are likely skewed in the same manner given the uniform construct of our spectral modeling.

\item On the other hand, we can point to the truncation of the disk in observation 3 as a potential explanation of the observed reduction of the reflected flux. If the disk recedes from the compact object (and by extension the illuminating corona assuming the corona is near the compact object and not the disk itself), we expect a reduction in the reflected flux as well. This does not account for the fact that there seems to be a negative correlation between the power-law flux and the flux of the reflected emission. However, as shown in Figure \ref{fig:plvsxillver}, for all but the highest power-law flux, the reflected flux is consistent within the 90\% uncertainty. 
\end{enumerate}

We can compare our results to those of \cite{koliopanos14}, who model the spectrum of \source using only an absorbed power-law plus a Gaussian to account for the bright \ion{O}{8} reflection feature. We find a slightly lower power-law index reported  than what is reported in this study, but our values are comparable ($\sim$~2 in our analysis as opposed to 2.3 in theirs). We may account for this difference by noting that \citet{koliopanos14} uses XMM-Newton, which is less sensitive in the harder X-ray regime and the models used therein do not account for low temperature thermal radiation, which could lead to a steeper power-law. Their model also does not account for the full reflection spectrum, and only includes a single feature to represent the reprocessed emission. This source was also characterized previously using \xillverco. \cite{madej14} find a value for the power-law index to be slightly higher than ours, again finding $\Gamma \approx 2.2$. They also find that both their disk and blackbody temperatures were slightly higher than ours. This could be due to the fact that \citeauthor{madej14} was using a different absorption component in their model. Their measured values of A$_{CO}$ are also significantly higher than the values we report. This is because at the time of their analysis, \citeauthor{madej14} was using an earlier version of \xillverco, that had a different initial chemical composition grid as discussed in \citet{ludlam21}. The remainder of the parameters in \xillverco are comparable with our results. In that study, they also model the spectrum of a different UCXB, 4U~1543$-$624. The measured values of the disk temperature for that source are very close to what is seen in our analysis of \source ($\sim 0.42$ keV), but their blackbody temperatures are significantly lower, indicating that these sources can be variable.

\section{Summary and Conclusions} \label{sec:conclusions}
 We have studied four simultaneous \nicer and \nustar observations of the UCXB \source, in an effort to track changes in the accretion disk as the system varies in flux. We modeled the spectra by combining a 3 component continuum model with a relativistically broadened reflection component from \textsc{xillverCO}. Our results can be summarized as follows. 
\begin{itemize}
    \item The flux of the reflected emission, as well as the flux of the thermal components representing emission from the accretion disk and the compact object itself, trend positively with the overall flux of the system.
    \item The flux of the power-law component, representing the emission from the corona, trends in the opposite direction. As the flux of the system decreases and reaches its lowest point, the emission from this power-law component is at its maximum. This is consistent with the low/hard state seen in BH LMXBs.
    \item During this low/hard state, we measure a slight disk truncation, with the inner disk being located close to 2~$R_{ISCO}$. This disk truncation is analogous to what is seen in BH LMXB systems, where below a certain Eddington ratio, the spectrum hardens and the disk recedes. We discuss other scenarios, but it is unclear with the current data set what the physical process driving this truncation is.
    \item We find that when the flux of the illuminating component (corona) is maximal, we see the minimum amount of reflected emission. A truncated disk can explain this discrepancy, since the inner disk is further from the illuminating corona. There may be some uncertainty in the contributions of the various components due to some model degeneracies, but the trends are consistent across all observations and with previous studies of similar sources.
\end{itemize}

Our data set only covers roughly half a period of the flux variation, so drawing conclusions about the long-term behavior of this source is difficult. Future observations are needed to completely understand what the driver of flux variation is, as well as understanding the truncation of the accretion disk and its relation to the spectral state of the source. Future missions like Athena \citep{nandra13}, or probe class concept missions such as HEX-P \citep{madsen19} or STROBE-X \citep{ray18}, could contribute immensely to such a study of the source by providing broad X-ray passband, larger effective collecting area, and/or higher energy resolution spectra. A deeper analysis of the faint Fe K$\alpha$ with higher energy resolution instruments could provide tighter constraints on the location of the inner disk, and could help to strengthen that both the Fe K and \ion{O}{8} features arise in the same location of the disk.
\\
\\
{\it Acknowledgements:} 
This work is supported by NASA under grant No. 80NSSC22K0054.


\begin{thebibliography}{}
\bibitem[Alabarta et al.(2020)]{alabarta20} Alabarta, K., Altamirano, D., M\'{e}ndez, M., et al.\ 2020, MNRAS, 497, 3896
\bibitem[Arnason et al.(2021)]{arnason21} Arnason, R.~M., Papei, H., Barmby, P., et al.\ 2021, \mnras, 502, 5455. doi:10.1093/mnras/stab345
\bibitem[Arnaud(1996)]{arnaud96}Arnaud, K. A.\ 1996, in Astronomical Society of the Pacific Conference Series, Vol. 101, Astronomical Data Analysis Software and Systems V, ed.\ G. H. Jacoby \& J. Barnes, 17
\bibitem[Bogdanov et al.(2019)]{bogdanov19}Bogdanov, S., Guillot, S., Ray, P. S., et al.\ 2019, ApJL, 887, L25
\bibitem[Brandt et al.(1992)]{brandt92} Brandt, S., Castro-Tirado, A.~J., Lund, N., et al.\ 1992, \aap, 262, L15
\bibitem[Cackett et al.(2008)]{cackett08}Cackett, E. M., Miller, J. M., Bhattacharyya, S., et al.\ 2008, ApJ, 674, 415
\bibitem[Cackett et al.(2009)]{cackett09b}Cackett, E. M., Altamirano, D., Patruno, A., et al.\ 2009b, \apjl, 694, L21
\bibitem[Cackett et al.(2010)]{cackett10}Cackett, E. M., Miller, J. M., Ballantyne, D. R., et al.\ 2010, ApJ, 720, 205
\bibitem[Chen et al.(2020)]{chen20} Chen, W.-C., Liu, D.-D., \& Wang, B.\ 2020, \apjl, 900, L8. doi:10.3847/2041-8213/abae66
\bibitem[Chiang et al.(2016)]{chiang16} Chiang, C.-Y., Morgan, R.~A., Cackett, E.~M., et al.\ 2016, \apj, 831, 45. doi:10.3847/0004-637X/831/1/45
\bibitem[Christian et al.(1994)]{christian94} Christian, D.~J., White, N.~E., \& Swank, J.~H.\ 1994, \apj, 422, 791. doi:10.1086/173771
\bibitem[Choudhury et al.(2017)]{choudhury17}Choudhury, K., Garc\'{i}a, J. A., Steiner, J. F., \& Bambi, C.\ 2017, ApJ, 851, 57
\bibitem[Churazov et al.(1996)]{churazov96} Churazov, E., Gilfanov, M., Forman, W., et al.\ 1996, \apj, 471, 673. doi:10.1086/177997
\bibitem[Coughenour et al.(2018)]{coughenour21} Coughenour, B.~M., Cackett, E.~M., Miller, J.~M., et al.\ 2018, \apj, 867, 64. doi:10.3847/1538-4357/aae098
\bibitem[Dauser et al.(2016)]{dauser16} Dauser, T., Garc{\'\i}a, J., Walton, D.~J., et al.\ 2016, \aap, 590, A76. doi:10.1051/0004-6361/201628135
\bibitem[Done et al.(2007)]{done07} Done, C., Gierli{\'n}ski, M., \& Kubota, A.\ 2007, \aapr, 15, 1. doi:10.1007/s00159-007-0006-1
\bibitem[Fabian et al.(1989)]{fabian89}Fabian, A. C., Rees, M. J., Stella, L., \& White, N. E.\ 1989, MNRAS, 238, 729
\bibitem[Frank et al.(2002)]{frank02} Frank, J., King, A., \& Raine, D.~J.\ 2002, Accretion Power in Astrophysics, by Juhan Frank and Andrew King and Derek Raine, pp. 398. ISBN 0521620538. Cambridge, UK: Cambridge University Press, February 2002., 398
\bibitem[Galloway et al.(2020)]{galloway20} Galloway, D.~K., in't Zand, J., Chenevez, J., et al.\ 2020, \apjs, 249, 32. doi:10.3847/1538-4365/ab9f2e
\bibitem[Garc{\'\i}a \& Kallman(2010)]{garcia10} Garc{\'\i}a, J. \& Kallman, T.~R.\ 2010, \apj, 718, 695. doi:10.1088/0004-637X/718/2/695
\bibitem[Garc\'{i}a et al.(2013)]{garcia13}Garc\'{i}a, J., Dauser, T., Reynolds, C. S., et al.\ 2013, ApJ, 768, 146
\bibitem[Gendreau et al.(2012)]{gendreau12}Gendreau, K. C., Arzoumanian, Z., \& Okajima, T.\ 2012, Proc. SPIE, 8443,13
\bibitem[Harrison et al.(2013)]{harrison13}Harrison, F. A., Craig, W. W., Christensen, F. E., et al.\ 2013, \apj, 770, 103
\bibitem[Ibragimov et al.(2005)]{ibragimov05} Ibragimov, A., Poutanen, J., Gilfanov, M., et al.\ 2005, \mnras, 362, 1435. doi:10.1111/j.1365-2966.2005.09415.x
\bibitem[Juett et al.(2001)]{juett01} Juett, A.~M., Psaltis, D., \& Chakrabarty, D.\ 2001, \apjl, 560, L59. doi:10.1086/324225
\bibitem[Kaastra \& Bleeker(2016)]{kaastra16} Kaastra, J.~S. \& Bleeker, J.~A.~M.\ 2016, \aap, 587, A151. doi:10.1051/0004-6361/201527395
\bibitem[King et al.(2016)]{king16} King, A.~L., Tomsick, J.~A., Miller, J.~M., et al.\ 2016, \apjl, 819, L29. doi:10.3847/2041-8205/819/2/L29
\bibitem[Koliopanos et al.(2014)]{koliopanos14} Koliopanos, F., Gilfanov, M., Bildsten, L., et al.\ 2014, The X-ray Universe 2014, 102. doi:10.48550/arXiv.1404.0617
\bibitem[Kuulkers et al.(2003)]{kuulkers03}Kuulkers, E., den Hartog, P.\ R., in’t Zand, J. J. M., et al.\ 2003, A\&A, 399, 663
\bibitem[Lazar et al.(2021)]{lazar21} Lazar, H., Tomsick, J.~A., Pike, S.~N., et al.\ 2021, \apj, 921, 155. doi:10.3847/1538-4357/ac1bab
\bibitem[Liu et al.(1999)]{liu99} Liu, B.~F., Yuan, W., Meyer, F., et al.\ 1999, \apjl, 527, L17. doi:10.1086/312383
\bibitem[Lodders(2003)]{lodders03}Lodders, K.\ 2003, ApJ, 591, 1220
\bibitem[Ludlam et al.(2019)]{ludlam19a}Ludlam, R.\ M., Miller, J.\ M., Barret, D., et al.\ 2019a, ApJ, 873, 99
\bibitem[Ludlam et al.(2020)]{ludlam20}Ludlam, R.\ M., Cackett, E.\ M., Garc\'{i}a, J.\ A., et al.\ 2020, ApJ, 895, 45
\bibitem[Ludlam et al.(2021)]{ludlam21} Ludlam, R.~M., Jaodand, A.~D., Garc{\'\i}a, J.~A., et al.\ 2021, \apj, 911, 123. doi:10.3847/1538-4357/abedb0
\bibitem[Ludlam et al.(2022)]{ludlam22} Ludlam, R.~M., Cackett, E.~M., Garc{\'\i}a, J.~A., et al.\ 2022, \apj, 927, 112
\bibitem[Madej et al.(2010)]{madej10} Madej, O.~K., Jonker, P.~G., Fabian, A.~C., et al.\ 2010, \mnras, 407, L11. doi:10.1111/j.1745-3933.2010.00892.x
\bibitem[Madej et al.(2014)]{madej14}Madej, O. K., Garc\'{i}a, J., Jonker, P. G., et al.\ 2014, MNRAS, 442, 1157
\bibitem[Madsen et al.(2019)]{madsen19} Madsen, K., Hickcox, R., Bachetti, M., et al.\ 2019, BAAS, 51, 166
\bibitem[Marino et al.(2020)]{marino20} Marino, A., Malzac, J., Del Santo, M., et al.\ 2020, \mnras, 498, 3351. doi:10.1093/mnras/staa2570
\bibitem[Migliari et al.(2010)]{migliari10} Migliari, S., Tomsick, J.~A., Miller-Jones, J.~C.~A., et al.\ 2010, \apj, 710, 117. doi:10.1088/0004-637X/710/1/117
\bibitem[Miller(2007)]{miller07}Miller, J. M.\ 2007, ARA\&A, 45, 441
\bibitem[Mondal et al.(2020)]{mondal20} Mondal, A.~S., Dewangan, G.~C., \& Raychaudhuri, B.\ 2020, \mnras, 494, 3177. doi:10.1093/mnras/staa1001
\bibitem[Nandra et al.(2013)]{nandra13}Nandra, K., Barret, D., Barcons, X., et al. 2013, arXiv:1306.2307
\bibitem[Nelemans et al.(2004)]{nelemans04}Nelemans, G., Jonker, P. G., Marsh, T. R., \& van der Klis, M.\ 2004, MNRAS, 348, L7
\bibitem[Nelemans et al.(2006)]{nelemans06} Nelemans, G., Jonker, P.~G., \& Steeghs, D.\ 2006, \mnras, 370, 255. doi:10.1111/j.1365-2966.2006.10496.x
\bibitem[Nelson et al.(1986)]{nelson86}Nelson, L. A., Rappaport, S. A., \& Joss, P. C.\ 1986, ApJ, 304, 231 
\bibitem[Ray et al.(2018)]{ray18}Ray, P. S., Arzoumanian, Z., Brandt, S., et al., 2018, Proc. SPIE, 10699, 19 (arXiv:1807.01179), doi:10.1117/12.2312257
\bibitem[Remillard et al.(2022)]{remillard22} Remillard, R. A., Loewenstein, M., Steiner, J. F., et al.\ 2022, AJ, 163, 130
\bibitem[Saavedra et al.(2023)]{saavedra23} Saavedra, E.~A., Garc{\'\i}a, F., Fogantini, F.~A., et al.\ 2023, \mnras. doi:10.1093/mnras/stad1157
\bibitem[Savonije et al.(1986)]{savonije86} Savonije, G. J., de Kool, M., \& van den Heuvel, E. P. J.\ 1986, A\&A, 155, 51
\bibitem[Shahbaz et al.(2008)]{shahbaz08} Shahbaz, T., Watson, C.~A., Zurita, C., et al.\ 2008, \pasp, 120, 848. doi:10.1086/590505
\bibitem[Shimura \& Takahara(1995)]{shimura95}Shimura, T., \& Takahara, R.\ 1995, ApJ, 445, 780
\bibitem[Steiner et al.(2010)]{steiner10} Steiner, J.~F., McClintock, J.~E., Remillard, R.~A., et al.\ 2010, \apjl, 718, L117. doi:10.1088/2041-8205/718/2/L117
\bibitem[Steiner et al.(2009)]{steiner09} Steiner, J. F., Narayan, R., McClintock, J. E., \& Ebisawa, K.\ 2009, PASP, 121, 1279
\bibitem[Syunyaev et al.(1991)]{syunyaev91} Syunyaev, R.~A., Arefev, V.~A., Borozdin, K.~N., et al.\ 1991, Soviet Astronomy Letters, 17, 409
\bibitem[Swank et al.(1978)]{swank78} Swank, J.~H., Becker, R.~H., Boldt, E.~A., et al.\ 1978, \mnras, 182, 349
\bibitem[Tomsick et al.(2009)]{tomsick09} Tomsick, J.~A., Yamaoka, K., Corbel, S., et al.\ 2009, \apjl, 707, L87. doi:10.1088/0004-637X/707/1/L87
\bibitem[van den Eijnden et al.(2020)]{eijnden20} van den Eijnden, J., Degenaar, N., Ludlam, R.~M., et al.\ 2020, \mnras, 493, 1318. doi:10.1093/mnras/staa423
\bibitem[Verner et al.(1996)]{verner96} Verner, D.~A., Ferland, G.~J., Korista, K.~T., et al.\ 1996, \apj, 465, 487. doi:10.1086/177435
\bibitem[Wilms et al.(2000)]{wilms00}Wilms, J., Allen, A., \& McCray, R.\ 2000, ApJ, 542, 914


\end{thebibliography}
\end{document}